\documentstyle[prl,aps,epsfig,floats]{revtex}

\let\footnote\savefootnote
 
\begin{document}
% \articletitle{
\title{Magnetic resonance peak and nonmagnetic impurities}

\author{Y. ~Sidis$^1$, P.~Bourges$^1$, B. Keimer$^{2}$, L. P. Regnault$^3$, J. Bossy$^4$, A. Ivanov$^5$, B.~Hennion$^1$,
P. ~Gautier-Picard$^1$, G.~Collin$^1$}

% \affil
\address{$^1$ Laboratoire L\'eon Brillouin, CEA-CNRS, CE-Saclay, 91191 Gif sur
Yvette, France.\\
$^2$ Max-Planck-Institut f\"ur Festk\"orperforschung, 70569 Stuttgart,
Germany.\\
$^3$ CEA Grenoble, D\'epartement de Recherche Fondamentale sur la Mati\`ere
Condens\'ee, \\
38054 Grenoble cedex 9, France.\\
$^4$ CNRS-CRTBT, BP 156, 38042 Grenoble cedex 9, France.\\
$^5$ Institut Laue Langevin, 156X, 38042 Grenoble cedex 9, France.\\
}
\maketitle

\begin{abstract}
Nonmagnetic Zn impurities are known to strongly suppress superconductivity.
We review their effects on the spin excitation spectrum in $\rm YBa_2Cu_3O_{7}$, as investigated by inelastic neutron scattering measurements.
\end{abstract}

\section{Introduction}
In optimally doped $\rm YBa_2Cu_3O_{6+x}$, the spin excitation spectrum is dominated 
by a sharp magnetic excitation at an energy of $\sim$40 meV and at the  planar antiferromagnetic
(AF) wave vector 
%$\rm {\bf q}_{AF}$
($\pi / a,\pi/ a$), the so-called magnetic resonance 
peak \cite{Rossat91,Mook93,Regnault94,Fong95,Bourges96,Regnault98}.
Its intensity decreases with increasing temperature and vanishes at
$\rm T_c$, without any significant shift of its characteristic energy
$\rm E_r$. In the underdoped regime,  $\rm E_r$ monotonically decreases  with decreasing hole concentration
\cite{Dai96,Fong97,Bourges97,Fong2000}, so that $\rm E_r \simeq$ 5 $\rm k_B T_c$.
Besides, it is possible to vary
$\rm T_c$ without changing the carrier concentration through impurity substitutions of Cu in the $\rm CuO_2$ planes. In $\rm YBa_2(Cu_{0.97}Ni_{0.03})_3O_7$ ($\rm T_c$=80 K), the magnetic resonance peak shifts to lower energy with a preserved $\rm E_r$/$\rm T_c$ ratio \cite{Sidis00}.

In optimally doped $\rm Bi_2Sr_2CaCu_2O_{8+\delta}$ ($\rm T_c$=91 K), a similar magnetic resonance peak has been recently observed at 43 meV \cite{Bi2212}. Furthermore, $\rm E_r$ shifts  down to 38 meV in the overdoped regime
 ($\rm T_c$=80 K) \cite{He2000}, preserving a constant ratio with $\rm T_c$: $\rm E_r  \simeq$  5.4 $\rm k_B T_c$.
Thus, whatever the hole doping, the energy position of the magnetic resonance peak always scales with $\rm T_c$.

In underdoped $\rm YBa_2Cu_3O_{6+x}$ (x=0.6,$\rm T_c$=63 K, $\rm E_r$=34 meV), recent INS measurements
provide evidence for incommensurate-like spin fluctuations at 24 meV and low temperature 
(seemingly similar to those observed $\rm La_{2-x}Sr_xCuO_4$) \cite{Dai-incom,Mook-incom}.
These incommensurate-like spin fluctuations are also observed at higher oxygen concentrations:
x=0.7
\cite{Bourges-Miami,Arai}, x=0.85 \cite{Bourges-science}. As a function of temperature \cite{Bourges-Miami,Bourges-science} and energy \cite{Bourges-science}, the incommensurability increases below $\rm T_c$ with decreasing temperature and decreases
upon approaching $\rm E_r$ in the superconducting state.
The results point towards an unified description of both incommensurate spin excitations and magnetic resonance peak in terms of  
an unique (dispersive)
collective spin excitation mode, as predicted in Ref.~\cite{Onufrieva}. 

In this paper, we review effects of nonmagnetic Zn impurities on the magnetic resonance peak in $\rm YBa_2Cu_3O_{7}$.
Among all candidates for substitution to Cu in the $\rm CuO_2$ planes of $\rm YBa_2Cu_3O_{7}$,
nonmagnetic Zn$^{2+}$ ions (3d$^{10}$, S=0)
induce the strongest $\rm T_c$ reduction ($\sim$ -12 K/$\%$ Zn) \cite{Tarascon88,Mendels99}.
Furthermore, low Zn substitution preserves the doping level and introduces only minimal structural
disorder.
We compare the spin excitation spectra reported from inelastic neutron scattering (INS) measurements performed in $\rm YBa_2(Cu_{1-y}Zn_{y})_3O_{7}$ for various Zn/Cu substitution rates \cite{Regnault98,Sidis00,Fong99,Sidis96,Sidis98} 
(Characteristics of single crystals used for INS measurements
are listed in Table \ref{ec}). Through Zn substitution, the magnetic resonance peak magnitude strongly decreases and its energy position slightly shifts to
lower energy, so that the ratio $\rm E_r$/$\rm T_c$ increases. In contrast to the Zn free system, where the normal state
 magnetic response is not experimentally discernible, nonmagnetic impurities
restore or enhance AF spin fluctuations above $\rm T_c$ and up to $\sim$250 K.

%In these review of the spin dynamics of $\rm YBa_2(Cu_{1-y}Zn{y})_3O_7$, we have focussed on the energy and temperature %dependence of $\rm \chi" (Q_{AF},E)$ within a restricted energy range around 40 meV. 

\section{INS measurements}

Throughout this review, the wave vector  $\rm {\bf Q}$
 is indexed in units of the reciprocal tetragonal lattice vectors $\rm 2\pi
/a=2\pi /b=1.63$ \AA$^{-1}$ and  $\rm 2\pi /c=0.53$ \AA$^{-1}$. In this notation
the $\rm (\pi/a,\pi/a)$ wave vector parallel to the ${\rm Cu O_2}$ planes
corresponds to points of the form (h/2,k/2) with h and k odd integers.
Because of the well known intensity modulation of the low energy spin
excitations due to interlayer interactions \cite{Rossat91,Mook93,Regnault94,Fong95,Bourges96,Regnault98}, 
data were taken close to L=1.7 $l$  
where $l$ is an odd integer.

\begin{figure}[t]
\vspace*{-0.1cm}
\epsfxsize=15cm
$$
\epsfbox{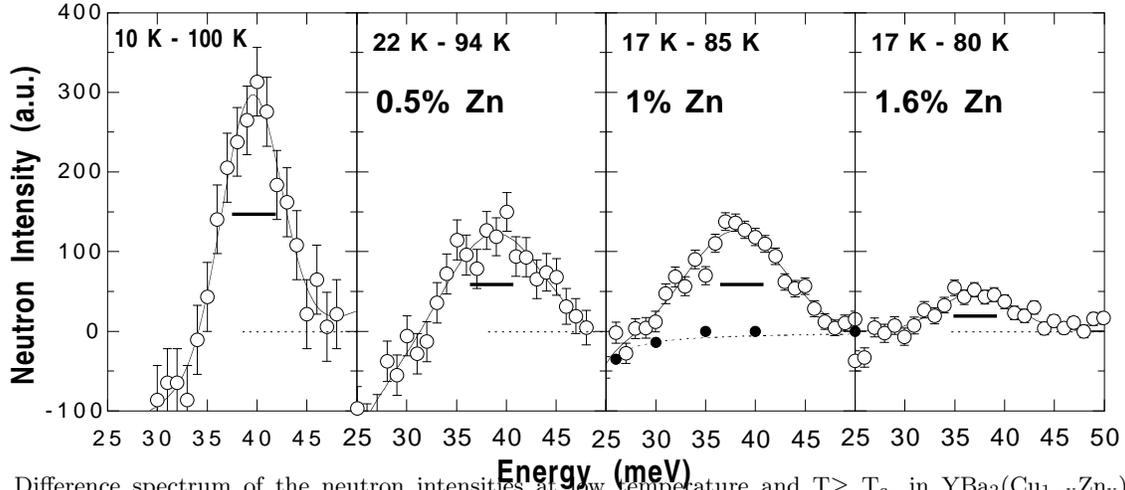}
$$
%\center{\epsfig{file=Zn-Escn.eps,height=40mm,angle=0}}
\vspace*{-1cm}
\caption[ff]{Difference spectrum of the neutron intensities 
at low temperature
and T$\ge$ $\rm T_c$, in $\rm YBa_2(Cu_{1-y}Zn_{y})_3O_7$ (open symbols):
a) y=0 \cite{Fong95} , b) y=0.005 \cite{Fong99}, c) y=0.01 \cite{Sidis00}, d) y=0.016. Measurements have been normalized against phonons.
Full symbols correspond to the reference level of magnetic scattering and is determined from the difference of constant-energy scans
at both temperatures. This level becomes slightly negative with decreasing energy  owing to  the thermal enhancement of the nuclear 
background. Dotted lines are guides to the eye. } 
\label{fig1}
\end{figure}

\begin{figure}[t]
\vspace*{-0.1cm}
\epsfxsize=15cm
$$
\epsfbox{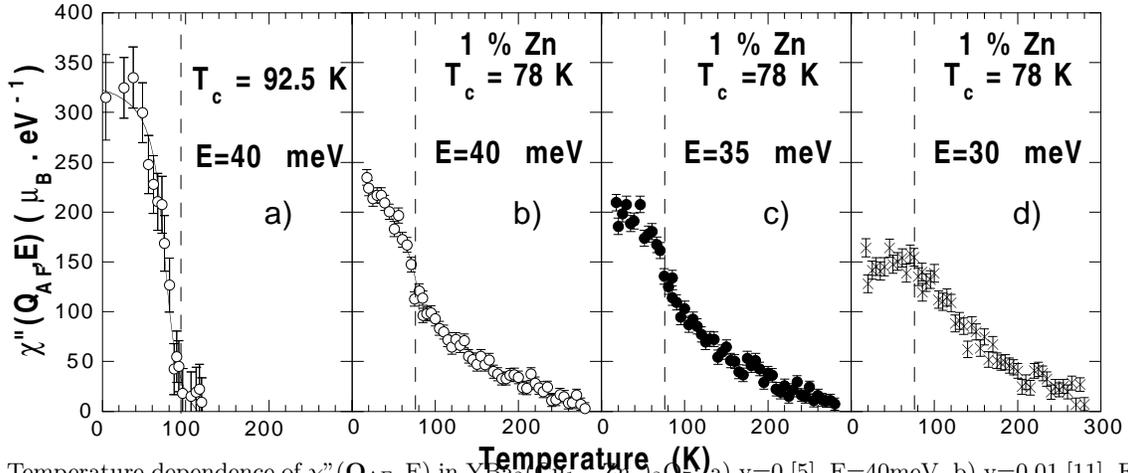}
$$
\vspace*{-1cm}
\caption[gg]{Temperature dependence of $\rm \chi "({\bf Q}_{AF}, E$) in
$\rm YBa_2(Cu_{1-y}Zn_{y})_3O_7$:
a) y=0 \cite{Bourges96}, E=40meV, b) y=0.01 \cite{Sidis00}, E=40 meV, E, c)y=0.01, E=35 meV, d) y=0.01, E=30 meV. Data are given in absolute units.
A $\sim$30$\%$ overall in absolute unit calibration is not included in 
the error bars. Solid lines are guides to the eye.}
\label{fig2}
\end{figure}

In pure $\rm YBa_2Cu_3O_{7}$, the magnetic resonance peak appears at $\rm E_r \sim$40 meV
\cite{Fong95}.
Figure \ref{fig1}.a shows the difference between two energy scans, performed with a momentum transfer fixed at $\rm {\bf Q}_{AF}$=(1.5,0.5,1.7). The former is measured deep in the superconducting state
%($\rm T \ll T_c$ )
 and the latter, in the normal state, close to $\rm T_c$.
%($\rm T \geq T_c K$).
The magnetic resonance peak gives rise to a positive contribution to the difference spectrum, with a maximum at
$\rm E_r$ (Fig.~\ref{fig1}.a). Besides, a negative difference at low energy stems from phonon scattering (determined independently through constant-energy scans).
The magnetic intensity has been converted to the imaginary part of the
dynamical magnetic susceptibility, $\chi"$, after correction by the detailed
balance and magnetic form factors and calibrated against optical phonons
according to a standard procedure \cite{Fong99}. The maximum at $\rm E_r$ in Fig.~\ref{fig1}.a then corresponds to an 
enhancement of $\chi"$ at the AF wave vector (hereafter $\rm \Delta \chi"({\bf Q}_{AF},E_r)$) of $\sim$ 300 $\rm \mu_B^2.eV^{-1}$. A fit to a Gaussian profile of
the positive part of the difference spectrum (Fig~\ref{fig1}) provides an estimate of the energy distribution of $\rm \Delta \chi"({\bf Q}_{AF},E)$ around $\rm E_r$. The full width at half maximum of the difference spectrum
is $\rm \Delta E  \sim$6 meV, of the same order of magnitude as the instrumental resolution ($\sim$5 meV), yielding an intrinsic energy width of at most $\sim$3 meV. In $\rm YBa_2Cu_3O_7$, the magnetic resonance peak is thus almost resolution limited in energy \cite{Regnault94,Fong95,Regnault98}.
The temperature dependence
of $ \rm \chi"({\bf Q}_{AF}$,40 meV), shown in Fig.~\ref{fig2}.a \cite{Bourges96}, exhibits a
marked change at T$_c$, and an order-parameter-like curve in the
superconducting state: the telltale signature of the resonance peak.
Above $\rm T_c$, magnetic fluctuations are not sizeable anymore. According to Ref.~\cite{Fong95,Bourges-Miami}, the magnitude 
of spin fluctuations left above $\rm T_c$ cannot exceed $\sim$70 $\rm \mu_B^2.eV^{-1}$. At 40 meV, the ratio, R,  between the intensities of AF spin fluctuations above $\rm T_c$ and at low temperature ranges from 0 to $\sim$ 20$\%$ (see Table.~\ref{ec}).

%figure 1 ----------------------------------

The same kind of  INS  measurements have been performed on Zn substituted $\rm 
YBa_2Cu_3O_7$ samples. The difference spectrum of neutron intensities for each Zn content is 
determined from energy scans performed at wave vector $\rm {\bf Q}_{AF}$=(-1.5,-0.5,1.7), following the procedure described above for the 
Zn free 
sample. For each Zn content, the difference spectra  still exhibit a positive maximum around 
$\sim$40 meV (Fig. \ref{fig1}), that accounts for an enhancement of the magnetic response at 
low temperature.
The intensity at the maximum drops down with increasing Zn 
substitution (Table ~\ref{ec}). Nonmagnetic impurities, that 
strongly reduce $\rm T_c$,  therefore significantly weaken the enhancement of the magnetic response in the superconducting state, ascribed to the magnetic 
resonance peak. The energy position of the maximum slightly moves to lower energy, giving rise to a progressive increase of the ratio $\rm E_r$/$\rm T_c$
from $\sim$ 5 to $\sim$ 6.
In addition, the energy distribution $\rm \Delta E$ ($\sim$8 meV) broadens with Zn 
substitution, as compared to the difference spectrum of the Zn free sample 
(Fig.~\ref{fig1}.a). The magnetic resonance peak thus exhibits an intrinsic
 energy width
that accounts for a disorder induced broadening \cite{Sidis00,Fong99}.

% As pointed out above, a minor
%reduction of the oxygen stoichiometry  may also contributed to this energy %broadening.

%figure 2 ----------------------------------

In $\rm YBa_2(Cu_{0.99}Zn_{0.01})_3O_7$,  the temperature dependence of
$\rm \chi" ({\bf Q}_{AF},40 meV )$ shows an upturn at $\rm T_c$  and displays
remnants of an order parameter lineshape 
in the superconducting state, that characterize the magnetic resonance peak
(Fig.~\ref{fig2}.b). The change of slope 
at $\rm T_c$ is hardly visible in the 0.5$\%$ and 2$\%$ Zn substituted samples
\cite{Regnault98,Fong99} (where data quality is not as high). Indeed, the hallmark 
of the magnetic resonance peak in the temperature dependence of $\rm \chi"$ is partially  
scrambled by AF spin fluctuations in the normal state
which are enhanced or restored by Zn.
These fluctuations  persist up to $\sim$250 K (Fig.~\ref{fig2}). Close to 40 meV, their relative weight 
with respect to the magnetic intensity at low temperature, R, increases with increasing 
Zn substitution (Table \ref{ec}).
Notice that the slight anomaly at $\rm T_c$ in 
the 1$\%$ Zn substituted sample is visible only because of the high quality of the data and that the improvement of data quality in other Zn substituted samples could reveal the same feature.

The temperature dependence of $\rm \chi" ({\bf Q}_{AF},E)$ has been measured at 
30 meV and 35 
meV in  $\rm YBa_2(Cu_{0.99}Zn_{0.01})_3O_7$ (Fig.~\ref{fig2}.c-d). In this 
system, the magnetic resonance peak appears precisely at 38 meV ($\rm \Delta E \simeq$8 meV) and  the 
enhancement of the magnetic response around $\rm T_c$ can be observed in the temperature 
dependences of $\rm \chi"$  at 35 meV and 40 meV (Fig.~\ref{fig2}.b-d). On the contrary,  $\rm \chi" ({\bf Q}_{AF}, 30 
meV)$ saturates or even slightly decreases below $\rm T_c$. A detailed
analysis of Fig.~\ref{fig2}.b-d reveals that the intensity of the magnetic response left in the normal state is actually larger at 30 and 35 meV than at 40 meV.
This implies a possible redistribution of the magnetic spectral weight
in the normal state.

%table ----------------------------------
\begin{table} [b]
\vspace*{-0.9cm}
%\mediumtext
\small
\center{
\begin{tabular}{ccccccc}
\hline
\multicolumn{1}{c}{y ($\%$)} &\multicolumn{1}{c}{V (cm$^3$)}
&\multicolumn{1}{c}{$\rm T_c$ (K)} &\multicolumn{1}{c}{$\rm E_r$ (meV)} 
&\multicolumn{1}{c}{$\rm \Delta \chi"$ ($\rm \mu_B^2.eV^{-1}$)} &\multicolumn{1}{c}{R ($\%$)}
&\multicolumn{1}{c}{Ref.}\\
%\tableline
\hline
0 & 10 &93 &40 &$\sim$300 &$\leq$20 &\cite{Fong95}\\
0.5 &1.7 &87 &39 &$\sim$130 &$\sim$50 &\cite{Fong99}\\
1 &$\sim$2 &78 &38 &$\sim$130 &$\sim$50 &\cite{Sidis00}\\
1.6 &$\sim$2 &73 &37 &$\sim$50 &- &-\\
2 &0.2 &69 &- &- &$\geq$70 &\cite{Sidis96,Regnault98}\\
\hline
\end{tabular}
\normalsize
}
%\vspace*{0.5cm}
\caption{$\rm YBa_2(Cu_{1-y}Zn_{y})_3O_{7}$ single crystals used in INS measurements: (a) Zn content,y, (b) volume,V, (c) superconducting critical temperature, $\rm T_c$. The samples were heat-treated to achieve full oxygenation
%\cite{Fong99}
and the Zn/Cu substitution rate was deduced from the reduction of T$_c$ as
compared to the pure system.
 In each sample, the magnetic resonance peak can be characterized by the following parameters: (d) the energy position, $\rm E_r$, (e) the enhancement, between T$\rm \rightarrow 0$ and $\rm T_c$, of dynamical spin susceptibility at $\rm E_r$ and $\rm Q_{AF}$, $\rm \Delta \chi"({\bf Q}_{AF},E_r)$. R corresponds  to the ratio between AF intensities left just above $\rm T_c$ and at low temperature: 
 R=$\rm \chi"({\bf Q}_{AF},E_r)_{T_c}$ / $\rm \chi"({\bf Q}_{AF},E_r)_{T \rightarrow 0}$. R is given at 39 or 40 meV. (for further details, see text)
 }
\label{ec}

\end{table}

\begin{figure}[t]
\vspace*{-0.1cm}
\epsfxsize=15cm
$$
\epsfbox{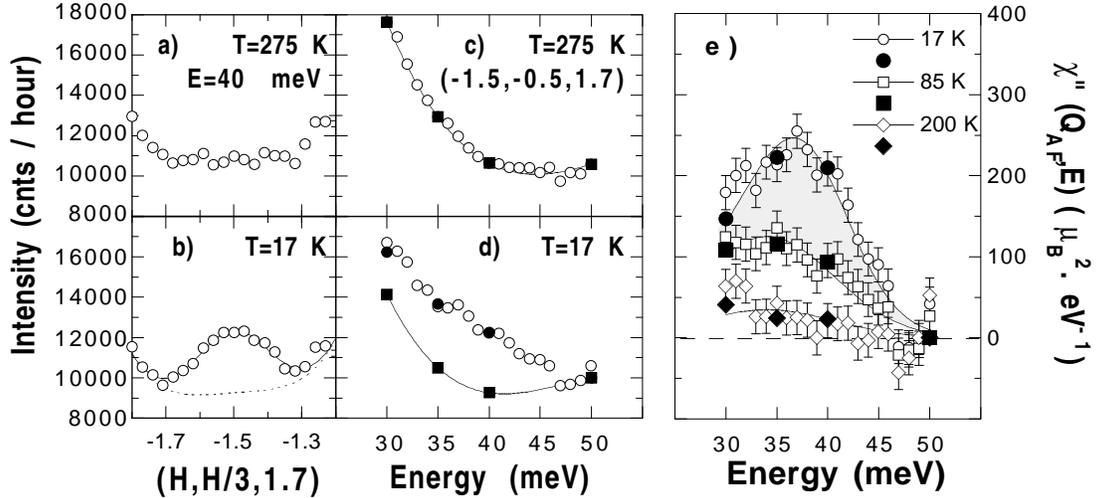}
$$
\vspace*{-1cm}
\caption{ YBa$_2$\-(Cu$_{0.99}$Zn$_{0.01}$)$_3$O$_{7- \delta}$. 
Constant energy scans at 40 meV in the (H,H/3,1.7) zone: a) 275 K, b) 17 K). Energy scan at the wave vector  (-1.5,-0.5,1.7):
c) 275 K, d) 17 K). Full circles and squares account for the magnetic intensity and background intensities determined from constant energy scans at different
energies. The lineshape of the background is fitted to a third order
polynomial function. e) $\rm \chi "({\bf Q}_{AF}, E)$ in the energy range 30-50 meV, at different temperatures. The shaded area corresponds to
$\rm \Delta \chi "({\bf Q}_{AF}, E)$ reported in Fig.\ref{fig1}.c.  Solid lines are guides to the eye.
%Full symbols correspond to the magnetic intensity determined from %constant-energy scans performed in the (3H,H,-1.7) zone.
%At room temperature, AF spin fluctuations vanish. 
%An energy scan performed at $\rm {\bf Q}_{AF}$ gives the energy dependence of %the background. It can be approximated by
%A linear interpolation between data at $\{ 30,35,40,50 \}$meV.
%At lower temperature, the interpolation between background intensities %determined from a set of constant-energy scans
%at $\{ 30,35,40,50 \}$meV defines an effective background. Its subtraction to %the raw energy scan provides a better definition of the magnetic excitation %spectrum lineshape(open symbols).
}
\label{fig3}
\end{figure}

Fig.~\ref{fig3}.a-b show constant-energy scans at 40 meV in the (H,H/3,1.7)
zone at 17 K and 275 K. At low temperature, the magnetic response displays
a Gaussian momentum distribution centered at the AF wave vector, on top of
a background that is slightly curved due to a contribution from phonons. At 275 K, the magnetic response is not sizeable anymore
(Fig.~\ref{fig3}.a) and an energy scan performed at $\rm {\bf Q}_{AF}$=(-1.5,-0.5,1.7) characterizes the energy dependence of the background
at high temperature (Fig.~\ref{fig3}.c). In the energy range E=30-50 meV, its
lineshape is well approximated by a third order polynomial fit of data at
$\{$30, 35, 40, 50$\}$ meV. At lower temperature, the same fit of background intensities determined from a set of constant-energy scans
at $\{ 30,35,40,50 \}$meV defines an effective background 
(Fig.~\ref{fig3}.d). Its subtraction from the raw intensity leads to the magnetic excitation spectrum (open symbols in Fig.~\ref{fig3}.e).
Figure \ref{fig3}.e shows  
the magnetic excitation spectrum at $\rm {\bf Q}_{AF}$ from 30 to 50 
meV in $\rm YBa_2(Cu_{0.99}Zn_{0.01})_3O_7$: in the superconducting state (17 K), close 
to $\rm T_c$ (85 K) and well above  $\rm T_c$ (200 K). In the normal state, the 
maximum of $\rm \chi" ({\bf Q}_{AF},E)$ moves inside the energy range 30-35 meV,  whereas the maximum intensity is still
peaked around $\sim$38 meV in the superconducting state.

%----------------------------- rejete dans caption figure 3.
%Following a standard procedure \cite{Fong99}, the nuclear and magnetic intensities at $\rm {\bf Q}_{AF}$
%were determined from constant-energy scans performed in the (3H,H,-1.7) in the 1$\%$ Zn substituted sample.
%The magnitude of AF spin fluctuations, put in absolute unit, is plotted 
%as function of energy in Fig~\ref{fig3}.b (full symbols).
%At room temperature, AF spin fluctuation do not contribute anymore and 
%an energy scan performed at the AF wave vector characterizes the energy dependence of the background.
%A linear interpolation between the intensities at $\{ 30,35,40,50 \}$meV approximate quite well energy scan.
%At lower temperature, the same interpolation between background intensities from a set of constant-energy scans
%at $\{ 30,35,40,50 \}$meV defined and effective background. Once this 
%effective background subtracted to the raw energy scan, a better definition of the magnetic excitation spectrum lineshape
%is obtained (open symbols in Fig.~\ref{fig3}.b).
%----------------------------- rejete dans caption figure 3.

We can summarize the experimental observations in $\rm YBa_2(Cu_{1-y}Zn_{y})_3O_7$ as follows.
In the superconducting state, the magnetic resonance peak broadens in energy and slightly moves to lower energy, but remains located close to $\sim$40 meV, the energy position of  
the resonance peak in pure ${\rm YBa_2 Cu_3 O_7}$. A broad peak
with a characteristic energy comparable to (but somewhat lower than) the
energy of the resonance peak appears in the normal-state response of Zn-substituted systems.
While the normal state AF spin fluctuations develop with increasing Zn substitution, the enhancement of 
the magnetic response, associated with magnetic resonance peak in the superconducting state, fades away.

%figure 3 ----------------------------------

\section{Discussion and conclusion}

A comparison of our measurements in the pure YBCO system with current models 
of the spin dynamics in cuprates
is given in \cite{Fong2000}. However, most of these models do not incorporate disorder. Here we restrict the discussion to theoretical works where
the interplay between (collective) spin excitations and quantum impurities in high temperature superconductors is expressly considered.

In BCS d-wave superconductors, nonmagnetic impurities cause the decay of the quasi-particle states due to a strong scattering rate (close to the unitary limit), and then, give rise to a pair breaking that reduces strongly 
the superconducting order parameter\cite{Hirschfeld86,Borkowski94,Hotta93,Sun95,Balatsky95,Kim95,Fehrenbacher96,Balatsky96,Franz96,Haran99}. The resonant scattering by non magnetic impurities qualitatively account for most of the Zn substitutions effects
in High-$\rm T_c$ superconductors:
i) the strong alteration the bulk superconducting properties, such as the
critical temperature \cite{Tarascon88,Mendels99} and the superfluid density \cite{Nachumi96,Bernhard96}, ii) the increase of the in-plane residual resistivity, iii) the reduction of the microwave surface resistance 
\cite{Bonn94} and 
iv) the appearance of a finite density of state at the Fermi level below $\rm T_c$ \cite{Ishida91}. Therefore, as a consequence of the reduction of the superconducting order parameter, the threshold of the electron hole spin 
flip continuum at the AF wave vector moves in principle to low energy as, 
in the superconducting 
state, that threshold energy is basically proportional to twice the maximum 
of the $d$-wave gap. Thus, in any models where the resonance appears at or below the continuum threshold, nonmagnetic impurities should lead to a shift of the resonance to lower energy and the occurrence of damping (so that no clear resonance is observed at large impurity concentrations) \cite{Li98,Morr98}.
These results  provide an explanation for the broadening of the magnetic resonance peak. Similar broadening can also occur due to disorder
in the paramagnetic state of quantum antiferromagnet\cite{Vojta99}.
However, the magnitude of the $\rm E_r$-renormalization
strongly depends on the model used to account for the magnetic resonance 
peak in the Zn-free system, and then, would be important to discriminate 
between the different models for the magnetic resonance peak.
Furthermore, the strong scattering by non magnetic impurities, so crucial in the superconducting state, also modifies the normal state properties.
In the normal state, the effect of nonmagnetic impurities on the the spin fluctuation spectral weight at the AF wave vector has been studied in the framework of the the 2D Hubbard model using the random phase approximation \cite{bulut99}. The main effect of dilute
impurities on the noninteracting dynamical spin susceptibility is a weak smearing. On the contrary, for an interacting system, the scattering of   
spin fluctuations by the (static and extended \cite{Ziegler96}) impurity potential with a finite momentum transfer ("umklapp"
processes) becomes essential. Indeed, the ($\pi/a,\pi/a)$ spin fluctuations 
become mixed with other wave vector components, and a new peak in 
$\rm \chi"({\bf Q}_{AF},\omega)$ can appear.

Scanning tunneling microscopy (STM) in Zn substituted $\rm Bi_2Sr_2CaCu_2O_{8+\delta}$ \cite{Pan00} confirms the existence of a strong quasi-particle scattering rate by impurities. Indeed, STM shows intense quasi-particle scattering resonances at Zn sites, coincident with strong suppression of the superconducting coherence peaks at the Zn site. Furthermore, the superconducting peaks then are progressively restored over a distance, $\xi$, of about 15 \AA. STM supports the proposal that the superfluid
density reduction can be explained by non-superconducting regions of area 
$\rm \sim \pi \xi ^2$ around each impurity atoms, the so-called
"Swiss cheese" model, introduced to account for the decrease of the superconducting condensate density from $\mu$SR measurements \cite{Nachumi96}.
% ($\xi$ is the superconducting correlation length, $\sim$20 \AA).
%Furthermore, the quasi-particle local density of states reveals
%a fourth fold symmetric quasi-particle cloud aligned with d-wave gap nodes.
On a phenomenological level, our
data are actually consistent with this scenario in which Zn impurities are surrounded by
extended regions 
%whose magnetic properties are strongly modified already far above ${\rm T_c}$, %and 
in which superconductivity never develops \cite{Nachumi96,Pan00}.
Within this picture,
superconductivity is then confined to (perhaps only rather narrow) regions
far from the Zn impurities. This would explain why Zn impurities all but
eradicate the effect of superconductivity on the spin excitations which is
so readily apparent in the pure system. Since INS is a bulk measurement and the magnetic resonance peak is an intrinsic feature of the superconducting state, one may speculate that its intensity may be suppressed as the fraction of the system that becomes superconducting and, thus, may scale with the superfluid condensate density. In addition, in non-superconducting regions around Zn impurities,  magnetic properties are strongly modified already far above ${\rm T_c}$. According to nuclear magnetic resonance measurements, local magnetic moments develop on Cu sites surrounding a 
Zn impurity (up to the third nearest neighbors)\cite{Mahajan94}.

The modification of the local magnetic properties and the resonant scattering from Zn impurities arise in a natural way from the strong 
correlation of the host
% and justify the use of the unitary limit.
as shown in exact diagonalizations of small clusters
performed on the framework of the t-J
 model \cite{Poilblanc94,Ohta96,Ziegler96,Odashima97,Riera96}. 
According to these calculations,  electrons form a bound state around the impurity site and the magnitude of the local moment is enhanced, as observed experimentally \cite{Mahajan94,Mendels99}.  When J/t becomes larger than $\sim$0.3, a mobile hole is trapped by the impurity potential induced by the local distortion of the AF background. Below $\rm T_c$, a pair breaking effect occurs due to the binding of holes to the impurity.
Likewise, a new magnetic excitation corresponding to the singlet-triplet excitation of the singlet impurity-hole bound state is predicted \cite{Riera96}.
The nucleation of staggered magnetic moment where superconductivity is suppressed and/or the singlet-triplet excitation of the singlet impurity-hole bound state \cite{Riera96}  may contribute to the
a broad peak observed in the normal state up to $\sim$250 K, with a characteristic energy smaller than $\rm E_r$ (but of the same order of magnitude). In this picture, the magnetic resonance peak
and spin fluctuations intrinsic to Cu spins surrounding Zn impurities coexist in the superconducting state.

%In \cite{Morr98,Vojta99}, the magnetic resonance peak corresponds to 
%a collective spin excitation mode, reminiscent of spin waves in the AF %insulating state.
%Whereas, in \cite{Morr98}, the magnetic resonance peak decays into electron %hole spin flip excitation through Zn substitution (as described above), the %scattering of the collective excitation by impurities is directly responsible %for the renormalization of its energy position and its damping in %\cite{Vojta99}.

In conclusion, our INS data show that the interplay between non magnetic quantum impurities and spin dynamics in the
cuprates is a surprisingly rich field of investigation, that
emphasized the importance of strong correlation and the competition between the superconducting ground state and antiferromagnetism. We hope that this review will stimulate further theoretical and experimental work.

\begin{acknowledgments}
The authors wish to acknowledge P. Hirschfeld, J. Bobroff, P. Pfeuty and F. Onufrieva for helpful discussions.
\end{acknowledgments}

% \begin{chapthebibliography}{99}

\end{document}